\documentclass[jnr,crcready]{iosart2x}
\usepackage{amssymb}
\usepackage{amsmath}
\usepackage{dcolumn}
\usepackage{multirow}

\pubyear{0000}
\volume{0}
\firstpage{1}
\lastpage{1}

\begin{document}

\begin{frontmatter}

\title{A Prototype Compact Accelerator-based Neutron Source (CANS) for Canada}


\author[A]{\inits{}\fnms{Robert} \snm{Laxdal}\ead[label=e1]{lax@triumf.ca}%
\thanks{Corresponding author. \printead{e1}.}},
\author[B,C]{\inits{}\fnms{Dalini} \snm{Maharaj}\ead[label=e2]{second@somewhere.com}},
\author[A,D]{\inits{}\fnms{Mina} \snm{Abbaslou}},
\author[E]{\inits{}\fnms{Zin} \snm{Tun}\ead[label=e2]{second@somewhere.com}},
\author[E]{\inits{}\fnms{Daniel} \snm{Banks}\ead[label=e2]{second@somewhere.com}},
\author[C]{\inits{}\fnms{Alexander} \snm{Gottberg}\ead[label=e2]{second@somewhere.com}},
\author[A]{\inits{}\fnms{Marco} \snm{Marchetto}\ead[label=e2]{second@somewhere.com}},
\author[A]{\inits{}\fnms{Eduardo} \snm{Rodriguez}\ead[label=e2]{second@somewhere.com}},
\author[F]{\inits{}\fnms{Zahra} \snm{Yamani}\ead[label=e2]{second@somewhere.com}},
\author[F]{\inits{}\fnms{Helmut} \snm{Fritzsche}\ead[label=e2]{second@somewhere.com}},
\author[F]{\inits{}\fnms{Ronald} \snm{Rogge}\ead[label=e2]{second@somewhere.com}},
\author[G,H,I]{\inits{}\fnms{Ming} \snm{Pan}\ead[label=e2]{second@somewhere.com}},
\author[A]{\inits{}\fnms{Oliver} \snm{Kester}\ead[label=e2]{okester@triumf.ca}%
\thanks{Corresponding author. \printead{e2}.}}
and 
\author[B,G]{\inits{}\fnms{Drew} \snm{Marquardt}\ead[label=e3]{drew.marquardt@uwindsor.ca}%
\thanks{Corresponding author. \printead{e3}.}}

\address[A]{Accelerator Division \orgname{TRIUMF},
BC, \cny{Canada}}
\address[B]{Department of Chemistry and Biochemistry, \orgname{University of Windsor},
ON, \cny{Canada}}
\address[C]{Targets \& Ion Sources \orgname{TRIUMF},
BC, \cny{Canada}}
\address[D]{Department of Physics, \orgname{University of Victoria},
BC, \cny{Canada}}
\address[E]{\orgname{TVB Associates Inc.},
ON, \cny{Canada}}
\address[F]{ \orgname{Canadian Nuclear Laboratories},
ON, \cny{Canada}}
\address[G]{Department of Physics, \orgname{University of Windsor},
 ON, \cny{Canada}}
\address[H]{Radation Oncology, \orgname{Windsor Regional Hospital},
ON, \cny{Canada}}
\address[I]{Schulich School of Medicine \& Dentistry, \orgname{Western University}, ON, \cny{Canada}}

\begin{abstract}
Canada’s access to neutron beams for neutron scattering was significantly curtailed in 2018 with the closure of the National Research Universal (NRU) reactor in Chalk River, Ontario, Canada. New sources are needed for the long-term; otherwise, access will only become harder as the global supply shrinks. Compact Accelerator-based Neutron Sources (CANS) offer the possibility of an intense source of neutrons with a capital cost significantly lower than spallation sources. In this paper, we propose a CANS for Canada. The proposal is staged with the first stage offering a medium neutron-flux, linac-based approach for neutron scattering that is also coupled with a boron neutron capture therapy (BNCT) station and a positron emission tomography (PET) isotope station. The first stage will serve as a prototype for a second stage: a higher brightness, higher cost facility that could be viewed as a national centre for neutron applications. 
\end{abstract}


\end{frontmatter}




\section{Introduction}
Most bright neutron sources are nuclear fission reactors that continuously produce massive quantities of neutrons at the source. Such reactors have stringent safety requirements and high costs for construction and operation. Spallation sources, such as the Spallation Neutron Source (SNS) \cite{Galambos2019} and the European Spallation Source (ESS) \cite{Garoby2017}, avoid many of the political and safety issues related to nuclear reactors by avoiding nuclear fuel. They typically have a lower time-averaged neutron flux, but by producing neutrons in pulses and using time-of-flight neutron beam methods, they can use the neutrons up to a factor of 100 more efficiently than continuous sources, which offsets the lower flux. For reactors and spallation sources of comparable performance at the instruments, spallation sources costs at least are in the same order of magnitude as reactors because they require a high-energy ($>$ 1 GeV) proton linear accelerator (“linac”). Spallation sources can achieve very high neutron brightness, but the requirement for GeV protons means they cannot be scaled down to achieve a lower-cost alternative to medium-brightness reactors.  

A Compact Accelerator-based Neutron Source (CANS) produces neutrons through a nuclear reaction between a low-energy (as low as a few MeV) proton beam and a beryllium or lithium target. A CANS can benefit from the same efficiency gains of a spallation source by producing neutrons in pulses and utilizing time-of-flight neutron beam methods. At these low energies, the cost of a linac and radiation safety requirements are much less than for spallation, and the footprint of a CANS is also much smaller. Specifically, it requires less shielding and infrastructure for cooling, thereby making the target-moderator assembly smaller (hence, “compact”), which allows for a neutron guide or an instrument to be positioned closer to the source in order to collect more of the neutrons for end uses. The efficiency of neutron production is orders of magnitude lower in a CANS than for spallation source.  For example the neutron yield for spallation reactions is about 25$-$30 neutrons per proton while $10^{-3}$ neutrons per proton are produced in the $^9$Be(p,n)$^9$B reaction, which is employed at a CANS \cite{Anderson2016}. But some performance loss can be compensated by the fact that experiments can be closer to the source. Further, the low cost of the linac and target-moderator assembly creates an option for multiple, independent sources (e.g. cold, thermal, and hot), each optimized for different end uses. 

Several CANS have been built for low-brightness neutron science, LENS \cite{Baxter2005} and RANS \cite{Otake2018} for example, and applications such as BNCT. The only CANS initiatives for neutron diffraction aiming higher than low-brightness applications are projects by the J\"{u}lich Centre for Neutron Sciences (JCNS) and the French Alternative Energies and Atomic Energy Commission (CEA) to develop designs for much brighter CANS to meet much of the needs of Germany and France, respectively. Their independent simulations suggest that such national-scale designs would achieve beamline performances that rival medium-brightness reactors at an expected price tag of C\$200M-300M~\cite{Gutberlet2019,Voigt2018}. Scaling-down such a source for Canada’s needs (e.g. ISIS-like brightness and 9 instruments) would be C\$100M-200M~\cite{Banks2019}. But it is unlikely that any nation would invest in such large infrastructure before the technology risk is mitigated through a smaller scale prototype.  

Thus, we propose to build a Prototype Canadian CANS (PC-CANS) that will demonstrate competitive neutron fluxes for diffraction and show that CANS technology offers a cost-effective alternative for medium-brightness sources. Our approach is patterned after the J\"{u}lich NOVA ERA concept \cite{Mauerhofer2017} that serves as a reference design for the J\"{u}lich High Brightness Neutron Source (HBS) project \cite{Gutberlet2019}. 
To attract users to PC-CANS, the prototype’s performance should be not less than a factor of 5 lower than the ISIS First Target Station \cite{Kockelmann2013}. NOVA ERA is a factor of about 25 down in brightness from ISIS (whether measured by time-averaged primary neutron generation rate at the source or by time-averaged neutron flux on the sample at their SANS instruments operating at the same resolution). Thus, our goal is to explore designs to improve brightness over NOVA ERA by a factor of five, without also increasing the cost by the same proportion. A grouping of neutron sources that serve as comparators for PC-CANS is given in Table~\ref{Tab_sources}.

\begin{figure}[htbp!]
    \centering
    \includegraphics[width=0.95\textwidth]{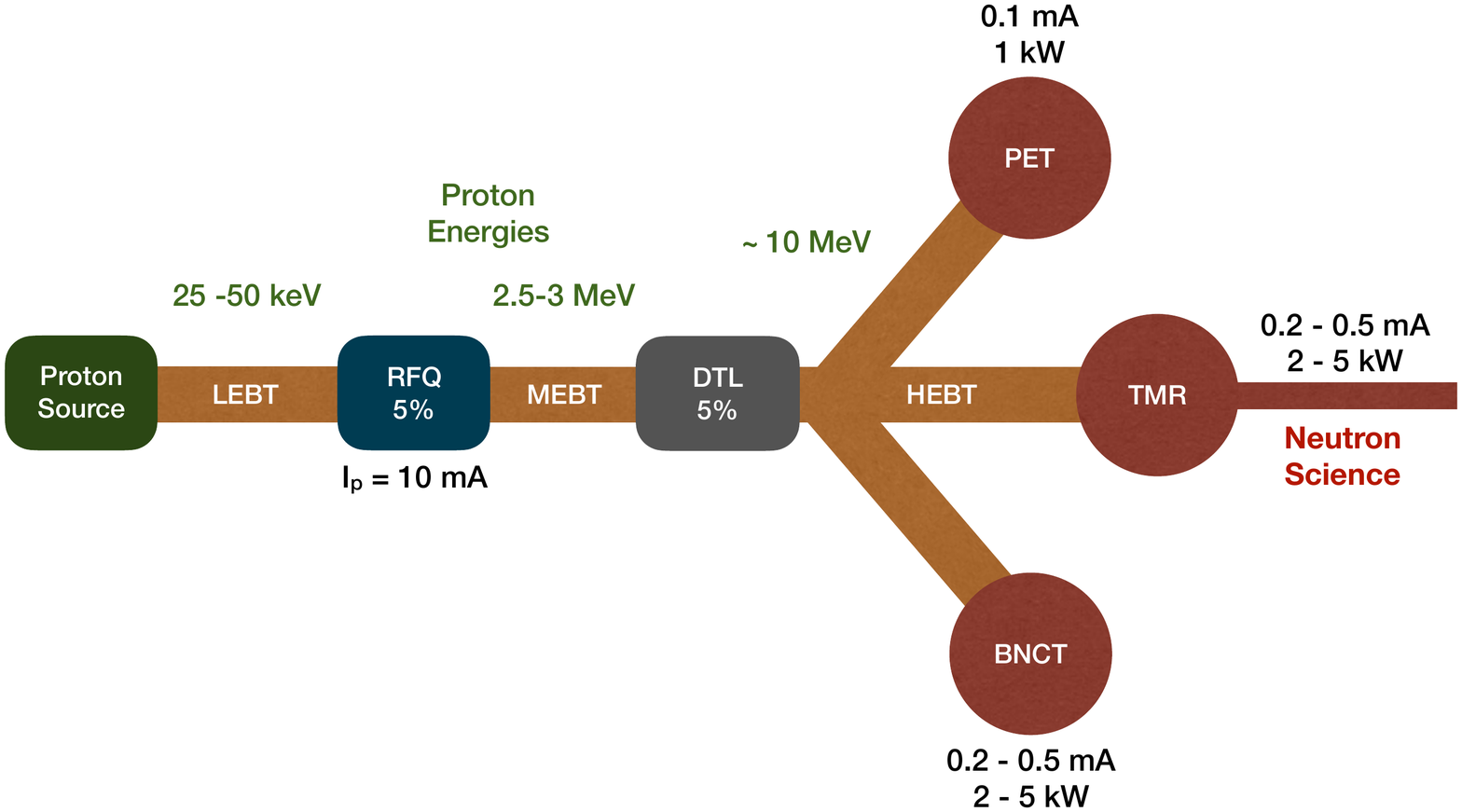}
    \caption{Schematic layout of PC-CANS}
    \label{fig:CANS_schematic}
\end{figure}

The entry level PC-CANS facility would be hosted at the University of Windsor and used to develop and demonstrate technology for a more powerful second stage, C-CANS (Canadian CANS) that would be a national-scale facility and could be located elsewhere. In addition to neutron source development and neutron scattering, the Windsor installation is also planned to enable BNCT research as well as $^{18}$F isotope production for PET. A schematic of PC-CANS is shown in Fig. \ref{fig:CANS_schematic} and the base requirements are discussed in the next section.

\section{CANS concept and requirements}
Our plan is to target a facility in the C\$10~M range for PC-CANS with a staging towards \$100M national facility, C-CANS (Table \ref{Tab_sources}). We would anticipate that the C-CANS would have a peak performance 5-10 times higher than PC-CANS. In terms of performance the goal is to design a PC-CANS with a production range that would give five times the useful neutrons as the NOVA proposal (especially for SANS). The NOVA parameters include a 10~MeV proton beam with a time-averaged beam intensity/power of 40 $\mu$A/0.4~kW, operating at a duty factor (DF) of 4\% for a peak intensity/power of 1~mA/10~kW with repetition rates from 48 to 384~Hz. The repetition rates and DF correspond to pulse lengths varying from 0.1 to 0.8~ms. A 100 $\mu$s pulse would result in thermal neutrons of <1ms duration after moderation. The NOVA proposal \cite{Mauerhofer2017} utilizes a water-cooled beryllium target for production of fast neutrons through the $^9Be(p,n)^9B$ nuclear reaction. At 10~MeV p on Be the expected yield is 2.1~$\times~10^{13}$ n/sec/mA. The neutron rate for the NOVA small angle neutron scattering (SANS) facility is expected to be up to 7~$\times$~10$^4$~n/cm$^2$/s. The most straightforward way to boost the neutron yield of NOVA by a factor of five is to increase the proton intensity by a factor of five keeping all other parameters equal. Thus the specification for PC-CANS on the Neutron Science target is a time-averaged beam intensity/power of 200 $\mu$A/2 kW. We choose to slightly increase the duty factor to 5\% for a peak intensity/power of 4 mA/40 kW with rep rates near 100~Hz. 

\begin{figure}[htbp!]
    \centering
    \includegraphics[width=0.95\textwidth]{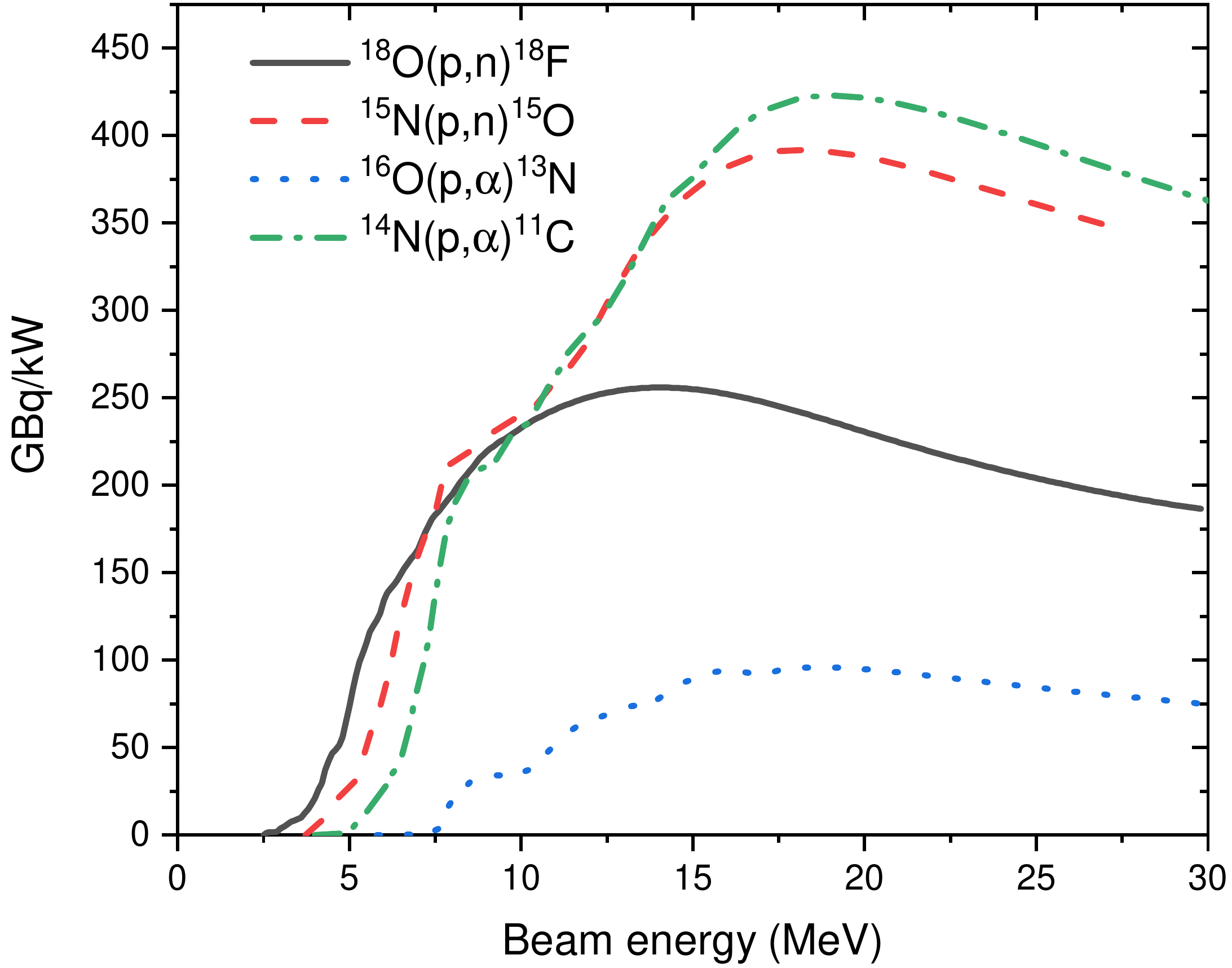}
    \caption{Yield of PET isotopes in dependence on the proton beam energy. Data accessed from the database in ref.\cite{Takacs2019}.}
    \label{fig:PET_isotopes}
\end{figure}

PET isotope production accelerators are typically operated from 10$-$18~MeV with lower energies putting more demand on the target window due to higher power loss. Competitive yields of $^{18}$F can be produced at 50$-$100$~\mu$A on target and the saturation yield of $^{18}$F as a function of beam energy is shown in Fig. \ref{fig:PET_isotopes}. Given the specification from neutron science considerations the specification for $^{18}$F end station calls for 10 MeV protons with a time-averaged current of 100~$\mu$A (1~kW on target) for a saturated yield of about 240~GBq. The time structure of the beam does not impact the yield while thermal cycling concerns would favour cw but is not an overriding design consideration.

The $^7Li(p,n)^7Be$ and $^9Be(p,n)^9B$ nuclear reactions are typically used for proton induced neutron production for BNCT. Neutron production on a Li target has a yield threshold of $\sim$1.9 MeV with a local production peak near 2.25 MeV. Typical accelerator based BNCT sources with Li targets operate with proton energies varying from 2.3$-$2.5 MeV. These low energy protons produce a lower yield than that achieved with higher energy protons, but the neutrons require less moderation to produce the therapeutic epithermal range for BNCT (0.5~eV~$<$~E$_{ep}$~$<$~10~keV) as stipulated by the International Atomic Energy Agency (IAEA) \cite{Agency2001}. For a dedicated BNCT facility, the use of low energy protons on a lithium target is cost-effective. However, lithium presents its own practical and engineering challenges given a high power proton source as it has a low melting temperature (180~$^\circ$C) and poor thermal conductivity (84.7~W$\cdot$m$^{-1}$K$^{-1}$), is highly reactive and produces radioactive $^7Be$ \cite{Jeon2020}. In addition to having much better thermo-mechanical properties than Li, with a melting temperature of 1287$^\circ$C and thermal conductivity of 190~W$\cdot$m$^{-1}$K$^{-1}$, bombardment of Be leads to stable products provided that the impinging proton energy is below 13.4 MeV \cite{Jeon2020}. These factors present significant advantages for target handling within a hospital-based environment. $^9$Be(p,n)$^9$B needs at least 4$-$5~MeV bombarding energy to produce a sufficient epithermal neutron yield. Given that the proton accelerator for PC-CANS has a peak energy of 10~MeV and neutron yields increase with energy, the BNCT target for PC-CANS will operate at 10~MeV and be made from beryllium. This choice has the advantage that the BNCT target development would leverage development on the Neutron Science target.  

The focus for BNCT research and development is to enable a staged approach to the eventual treatment of deep-seated tumors which require an epithermal neutron flux at the patient larger than 1.0~$\times$~10$^9$~n/cm$^2$/s~\cite{Agency2001}. Studies have found that at 10~MeV proton energy on a Be target with a moderator of MgF$_2$ gives an optimal epithermal neutron yield of 0.5~$\times$~10$^9$ n/s/cm$^2$/mA~\cite{Kiyanagi2019}. This is five times higher than the epithermal yield from 2.5~MeV protons on Li of 1E8~n/s/cm$^2$/mA~\cite{Zhu2018}. In conclusion to reach sufficient flux for a treatment facility a time-averaged current/power of 2~mA/20~kW would be required at 10~MeV.  

What does this mean for the PC-CANS BNCT station? Given that the main focus of PC-CANS is neutrons science, an accelerator based BNCT end-station will not drive the beam requirements. Instead, the BNCT station will be an entry level facility for pre-clinical research and would provide a development path for a future BNCT therapy machine. As reasoned above, our strategy is to utilize Be technology for both BNCT and the neutron science target. This allows that the target development for both neutron science and BNCT can proceed together. A BNCT facility with a time-averaged current of 200~$\mu$A at 10~MeV (2~kW) at 5\% duty factor would provide a peak current of 4~mA/40~kW and an epithermal neutron yield of 1.0~$\times$~10$^8$~n/s/cm$^2$ (10 times less than required for patient treatments of deep-seated tumours). This still presents a reasonable rate to perform research and development for BNCT technology and treatment.

As a strategy towards a staged upgrade, the linac is specified so that all end stations could be supplied with the required current simultaneously corresponding to a total average current across all end stations of 500~$\mu$A. This feature leaves open the option that more current could be supplied to a particular station with dedicated delivery should the evolution of target engineering allow an increase in beam intensity to produce a higher yield.

\section{Accelerator conceptual design}
\subsection{Overview}
The beauty of the CANS principle is that the performance level can vary from a low power pulsed neutron source designed for universities and industry with an average power at the target of around 1 kW to a high performance neutron source with tens of kWs in average power designed as a destination facility. Here we define an entry level CANS, the so called PC-CANS, with the flexibility to deliver neutron science but also to deliver protons to a BNCT development facility while producing $^{18}$F for PET.  

A typical layout of an accelerator-based neutron source, based on the global landscape, includes an ion injector, a linear accelerator (linac), a macro-pulse chopper (depending on the number of instruments), a Target-Moderator-Reflector-System (TMRS) that optimizes thermal neutron yield and minimizes background. The typical linac would include an RF quadrupole (RFQ) to accelerate from source potential to a few MeV followed by a drift tube linac (DTL) to reach the final energy. There are other accelerators that can deliver a fixed energy in the 10~MeV range, for example an electrostatic tandem or an industrial cyclotron. A cyclotron offers a cost-effective way to get to 10~MeV with cw beams suitable for BNCT and $^{18}$F production, but is limited in peak beam intensity to <1 mA and so precludes staging, while externally chopped beams for neutron time of flight (TOF) methods would lack intensity. The tandem accelerator is compatible with both pulsed and cw beams, but intensities would be limited by stripper foil life-time. A linac can achieve both high peak beam intensities and high time-averaged currents while providing staging opportunities for future upgrades. Room temperature linacs would operate in a pulsed regime that can be challenging for target design due to thermal cycling, but simplifies the linac engineering over cw variants and delivers high flux neutrons over a macro-pulse for applications that favour a high-brightness, pulsed neutron source.

In the case of the CANS, the ion source and low energy transport should be able to inject protons into the RFQ with a macro-pulse structure, that may be modified by a chopper, to meet the needs of the downstream target-instrument system.  The RFQ will take the low energy protons and accelerate them to an energy range of between 2.5-3~MeV, to an energy suitable to be injected into a DTL to reach higher energies. The rf power in the linac would be pulsed at the same macro-cycle to reduce the average power dissipated in the rf structure. The actual macro-structure would depend on the number of target stations and the requirements of the generated neutrons. For typical moderator systems, a 100~$\mu$sec proton pulse would generate a 1~msec pulse of thermal neutrons and so repetition rates of the order of a few hundred Hz would give sufficient resolution between thermal neutron pulses to allow TOF techniques. An option is to include a macro-pulse beam that can be fed to a multiplexer, where each micropulse is delivered to its respective end station. This provides the advantage of being able to support several TMRS systems with several instruments attached to each target but with only one linac.   

\subsection{PC-CANS Requirements}
The requirements of the prototype PC-CANS system call not only for a competitive neutron flux in a SANS facility, but also to have the capability to produce $^{18}$F for PET and an end station for BNCT development. As discussed in Section 2, $^{18}$F production cross sections favour higher proton bombardment energies >=10~MeV. BNCT facilities with Li targets have been designed for low energy making it possible to host the BNCT station after the RFQ, but target considerations and yield stated above have led us to choose BNCT production at 10~MeV on Be. 

As for the neutron science production target, even though lithium would aid in reducing the necessary beam energy to produce neutrons, high power lithium targets are more complex relative to beryllium targets, since, as the beam power increases, a liquid lithium target would be required that represents a major technical challenge. The increased cost and research and development required to implement a liquid lithium target, relative to a solid Beryllium target, make liquid target technology a better candidate for a future upgrade towards a high-end CANS facility.  

The PC-CANS layout consists of a high (peak) intensity proton linac with a source, an RFQ and a DTL delivering 10~MeV protons to three end stations: TMRS-Neutron Science (SANS and multi-purpose), TMRS-BNCT and a $^{18}$F production system. Each of the TMRS stations would be specified for average currents of 200~$\mu$A while the $^{18}$F station would be specified for a time-averaged current of 100~$\mu$A. At 5\% duty factor these correspond to peak intensities of 4~mA, 4~mA and 2~mA respectively. The peak intensity from the linac would be specified to allow achieving these beam intensities assuming a simultaneous time share between the end stations, so that a peak intensity of 10~mA at 5\% duty factor is foreseen.  Beyond this, to add further upgrade potential and engineering margin, the linac would be designed for twice this current or a maximum peak intensity of 20~mA corresponding to a peak power of 200~kW. The values are summarized in the Table~\ref{Tab_summary}.

In terms of performance, a time-averaged proton current of 200~$\mu$A would produce five times the average neutron flux of the NOVA ERA proposal. As target technology evolves the full proton intensity, 500~$\mu$A, could be sent to a single station for a time-averaged yield of 12.5 times NOVA ERA. Finally, for engineering margin the linac design would be compatible with a total average current of 1~mA such that, if the target engineering evolves, an average neutron yield of 25 times NOVA ERA could be reached. Since the protons are pulsed at 5\% duty factor the peak neutron flux is a factor of 20 higher than from the same cw machine demonstrating the key advantage over a 1~mA cw cyclotron. 

\subsection{Linac building blocks}
The pulsed low duty cycle scheme for the PC-CANS favors a normal conducting linac. In general, normal conducting cavities are more cost favourable than superconducting technology for low energy, high beam current and low duty cycle applications especially for a non-accelerator laboratory setting. The typical layout of a modern proton linac comprises an Electron Cyclotron Resonance Ion Source (ECRIS), a Low Beam Energy Transport section (LEBT), a Radio Frequency Quadrupole (RFQ), a Medium Energy Transport section (MEBT), a Drift Tube Linac (DTL) and a high energy beam transport (HEBT). Recent years have seen a significant progress in the development of a new generation of optimized linac systems and structures supported by large scale projects like FAIR \cite{Clemente2011}, ESS \cite{Eshraqi2014} and SPIRAL2 \cite{Ferdinand2013} for instance. The main considerations in choosing the technology are capital and operating costs, technical risk, reliability and availability. 

\subsection{Proton Source}
An Electron Cyclotron Resonance (ECR) ion source is the most favorable type of all available sources, providing high beam intensity continuously. Due to the microwave driven RF-discharge and the efficient heating of the source plasma by the ECR condition, high plasma densities and electron temperatures can be achieved. This is ideal for high intensity multi-mA proton beam production. The advantages in comparison to other proton source types like Penning Ion Gauge (PIG) sources, plasmatron sources or multi-cusp bucket sources are the high reliability and availability due to the simple structure (no filament), the easy handling and maintenance and a high proton fraction - there are fewer molecular ions, due to break up in the hot plasma. A source potential of 50~kV is typical for moderate space charge applications like PC-CANS. 

\subsection{LEBT}
High intensity proton low energy beam transport (LEBT) systems are typically compact with either electrostatic or magnetic elements, as well as the possibility for beam diagnostics. Here, we favour a reasonably compact LEBT but containing two solenoids for transverse focus and a short diagnostic box to help characterize the beam before RFQ acceleration. The initial beam chopping will be done in the LEBT as well to establish the macro-pulse synchronized to the rf macro-cycle.

\subsection{RFQ}
Acceleration of low energy protons is typically done in an RFQ. The RFQ is a low-velocity, high-current linear accelerator with high capture efficiency for hadron acceleration. The RFQ gives compact acceleration of good beam quality up to high beam intensities. To develop large voltages at the tips of the vanes at high frequencies requires resonant circuits. The two most-frequently used cavity structures are the four-vane structure, based on a ridged waveguide, and the four-rod structure, based on resonant quarter-wave supports that provide voltage to the 4 vanes. For both types of structure, the design process involves two largely orthogonal efforts: (1) Beam dynamics design resulting in parameters describing the RFQ electrode modulations using a simulation code, and (2) RF and mechanical design of the resonant cavity. 

The 4-vane structure operates in the TE120 mode and is the most popular for proton acceleration because its most common operating frequency range (150 to 500 MHz) is suitable for light ions where the injection velocity is relatively high. One issue is that when loaded by the vanes, the dipole and quadrupole modes can become degenerate and mix, producing a strong dipole field on the axis that needs care to avoid. Also, fabrication requires demanding specifications in terms of final dimensions and the structure is closed such that final access is restricted. The 4-rod structure has smaller transverse dimensions than a 4-vane RFQ at the same frequency, so that the 4-rod RFQ is often preferred for heavy ion applications where lower frequencies (<200~MHz) are applicable and the 4-vane structure would have a too large cross-section.  Advantages of the 4-rod are that the vacuum tank can be a lidded structure so that final access is straightforward and the even and odd rods are tied together such that the dipole mode is not supported. However for higher frequencies (above 200MHz) the dimensions are small leading to high power densities and the vertical asymmetry in the rf structure produces a field asymmetry on axis that needs compensation. For these reasons, RFQs operating at frequencies above 200 MHz and at a high duty cycle are usually of the four-vane type. 

The PC-CANS parameters allow some flexibility in the technical choice as the space charge forces are not extreme and the duty factor at 5\% considerably reduces rf power density in the structure. There are several high intensity pulsed proton RFQs operating in the energy range from 2.5-3~MeV. Examples include CERN Linac 4 \cite{Rossi2012}, ESS-Bilbao \cite{Munoz2017}, SNS \cite{Aleksandrov2017}, and CPHS \cite{Wang2014} RFQs. All have frequencies in the range from 300-400~MHz, a vane voltage of ~80~kV and a length of 3-4~m and employ 4-Vane RFQs. 

\subsection{MEBT}
The Medium Energy Beam Transport (MEBT) system matches the beam from the RFQ to the DTL. An important aspect of the beam transport to the DTL is the beam energy of the RFQ. This should be high enough to enable efficient beam transport along the MEBT considering space charge and longitudinal debunching and that the drift tube gap to gap distances set by rf frequency and particle velocity are reasonable. For a current near 20~mA, energy values between 2.5 and 3~MeV are a good choice. Transverse matching would be done by magnetic quadrupoles and longitudinal matching by a two-gap bunching cavity. The requirements of the transverse matching would depend on the DTL technology chosen.
An additional beam chopping system could also be added to impart flexibility to the beam delivery for customized time structures. 

\subsection{DTL}

The DTL will take the beam from the RFQ energy of 2.5-3~MeV and accelerate to 10~MeV. There are two typical technologies used in this energy range: an Alvarez DTL or an H-mode DTL.  In the case of protons where incoming velocities are relatively high, the H-mode structure of choice is the cross bar (CH)-DTL. The Alvarez operates in TM010 mode with 2$\pi$ phase shift between accelerating gaps. The drift tubes host either quadrupole electro-magnets or permanent magnets to provide strong transverse focus during acceleration typically in a FODO lattice. The CH-DTL utilizes the TE210 mode that can be used to generate on axis accelerating fields through cross-bar stems connected to drift tubes on the beam axis excited in $\pi$-mode. The H-mode structures are characterized by small drift tubes with no transverse focusing yielding a higher shunt impedance compared to the Alvarez but with the necessity that the accelerating sections are shorter with quadrupole triplets in between the tanks or incorporated in the tanks periodically. 

In cw mode the gradients for each are kept modest to reduce the dissipated rf power per meter while at duty factors considered for PC-CANS (5\%) the gradient can be chosen more aggressively (3-4~MV/m) dominated by the peak surface field on the drift tubes and considerations of the specification of the peak power required from the rf amplifier. Shunt impedance values for the Alvarez DTL operated with permanent magnet quadrupoles are expected to be 40~M$\Omega$/m. A single tank of 2.5~m will be required with an average power of $\sim$30~kW. Expected shunt impedance values for the CH structure in this energy range vary from  $\sim$80-110~M$\Omega$/m so rf power consumption would be less. An example layout consists of two sets of quadrupole triplets separating three accelerating sections to maintain transverse focusing with the overall length expected at $\sim$3~m. 

\subsection{HEBT}
The High Energy Beam Transport (HEBT) consists of magnetic quadrupoles to capture the diverging beam after acceleration and allow flexibility to tune the beam size on target. Depending on the target size a set of time varying steering magnets could also be added upstream of the targets to allow rastering over a prescribed beam size to maximize neutron yield while controlling the power density on the target. The HEBT will also be designed with a switch-yard to selectively allow the protons to feed any of the three end stations. A multiplexer feature with pulsed magnetic elements is also foreseen to be able to supply 2-3 end stations simultaneously. 

\subsection{Linac Summary}
A reasonable working rf frequency for the linac is in the range 320-360~MHz where several linacs have been built. This frequency range is particularly suited for a proton linear accelerator because it offers a good compromise between the large dimensions and relaxed fabrication tolerances of structures at low frequency and the high gradients and efficiencies allowed by higher frequencies. The envisaged linac would have accelerating sections (RFQ and DTL) within 6-7~m in length and the source, LEBT and MEBT would span 3~m for a total accelerator length of <10~m. 

A significant component of the linac cost is the capital cost of the rf power system so careful design/cost optimization studies are required to optimize rf efficiency. The beam loaded power required for accelerating 1~mA average (20mA~peak) would be 10~kW (200~kW) with the rf structures expected to dissipate 50~kW (1~MW). In this frequency and power range one or two rf klystrons with associated waveguide systems could supply the rf power. 

The final choice of the linac technology for the RFQ and DTL will be made later after considering capital costs and operating costs, reliability and operational complexity.

\section{Neutron production, moderation, and extraction}
\subsection{Overview}



The neutron generation process envisioned for PC-CANS is based on the approach outlined for the NOVA ERA, where a 10~MeV proton beam will be utilized in combination with a thin Be target thermally anchored to a water cooled backing conductor. Using target design principles previously developed for the TRIUMF-ARIEL facility\cite{Egoriti2020}, PC-CANS is aiming at a significantly higher
time-averaged power on target of 2~kW versus 0.4~kW in the case of NOVA ERA. Thermo-mechanical simulations\cite{Egoriti2020} as well as experimental irradiation tests using low energy electron beams\cite{Cade2018} show that power densities in excess of 10 W/mm$^2$ can be dissipated. However, we expect that hydrogen-induced blistering inside the Be layer or the material interface to the water-cooled substrate will be the lifetime limiting factor for the PC-CANS target design. For PC-CANS we hence aim at 2~kW time-averaged power with 30~mm diameter beam spot on a water-cooled 40~mm diameter Be target. Initial performance simulations utilize the NOVA ERA source geometry as a starting point for the PC-CANS target geometry, including the thickness of the moderator (polyethylene), and that of the neutron reflector (Pb).   

\subsection{Simulation Comparisons with Fluka and MCNP6}

\begin{figure}[htbp!]
    \centering
    \includegraphics[width=0.9\textwidth]{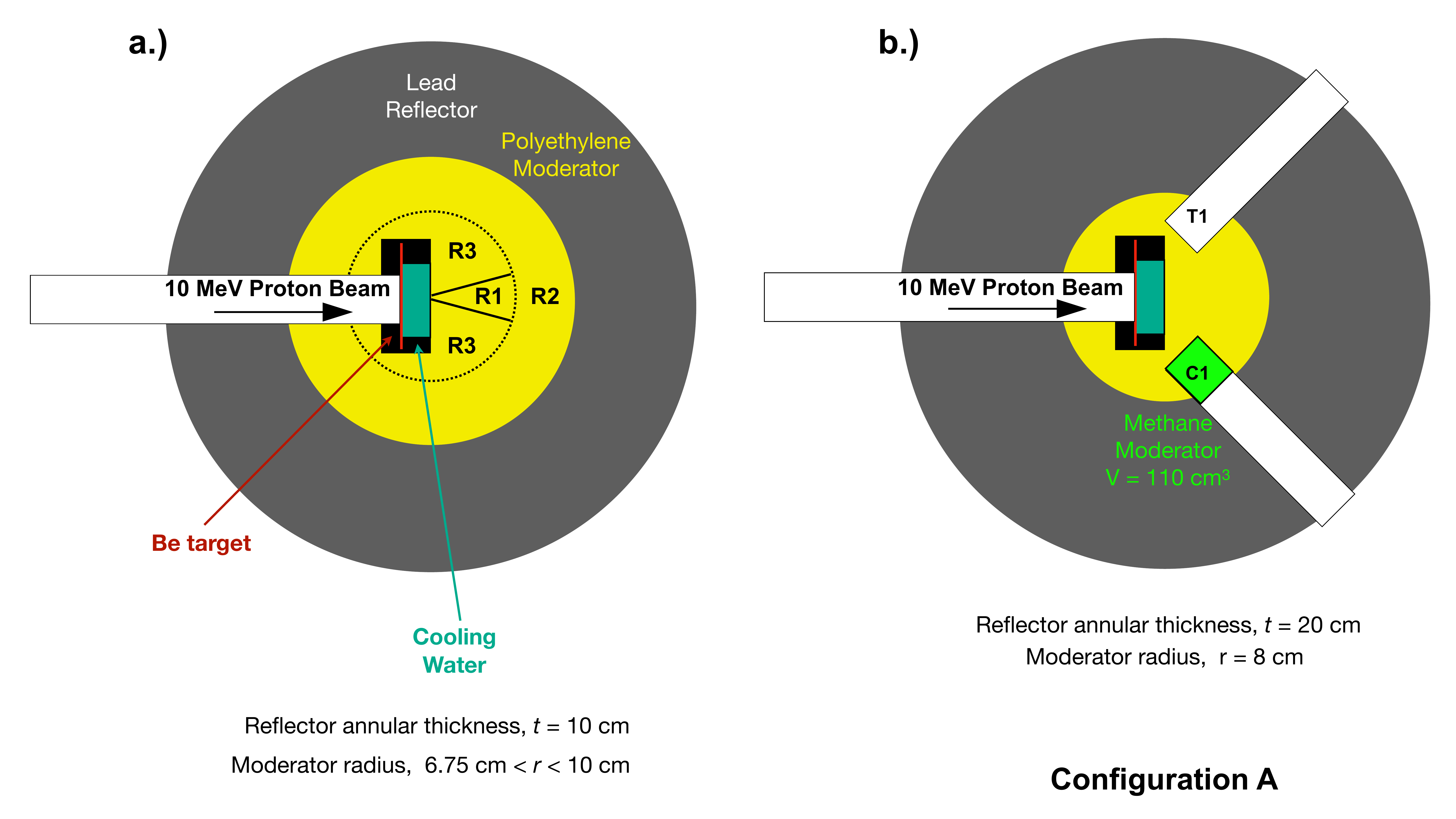}
    \caption{The geometry utilized in simulation comparisons are illustrated in this figure. Configuration A in panel b.) refers to a geometry which is utilized in the NOVA ERA report, shown on page 25 of Reference \cite{Mauerhofer2017}.}
    \label{fig:CombGeo}
\end{figure}

\begin{figure}[htbp!]
    \centering
    \includegraphics[width=0.7\textwidth]{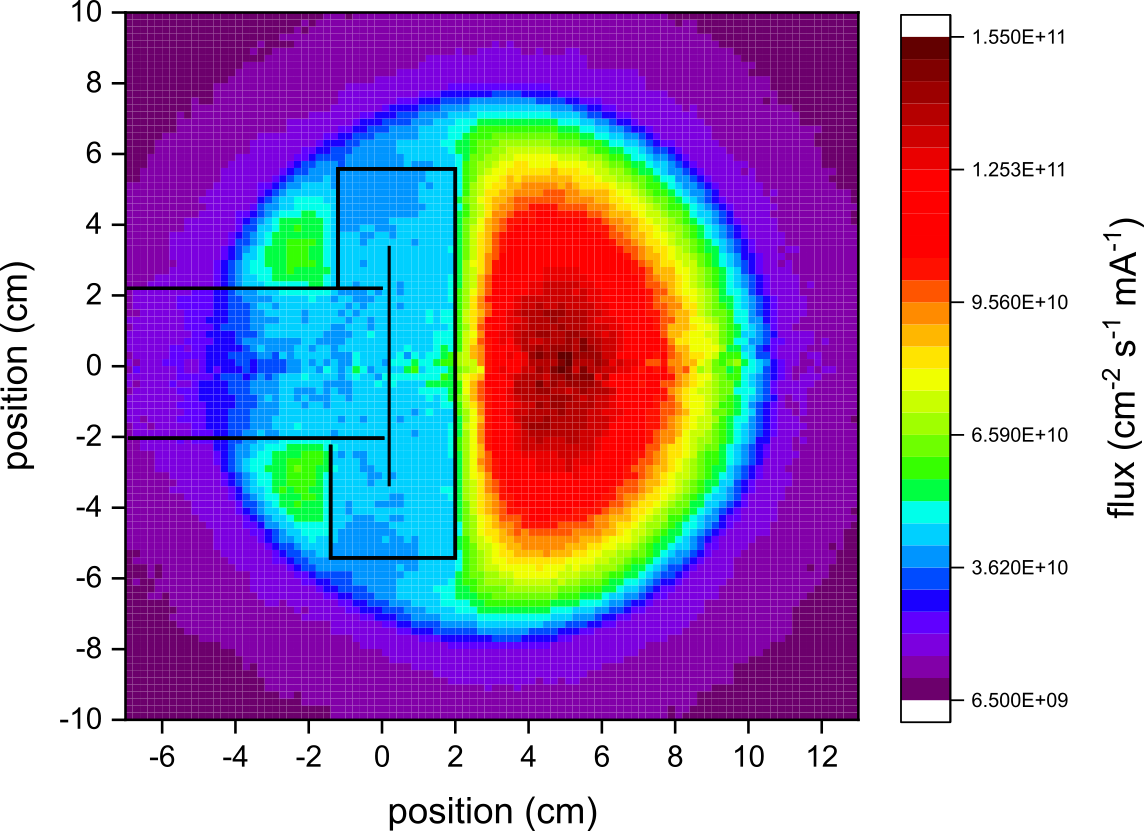}
    \caption{Spatial distribution of unperturbed thermal neutron flux within a polyethylene moderator and lead reflector generated from a beryllium target irradiated with 10 MeV protons. The thermal neutron flux includes the integrated neutron spectra. The geometry in Fig. \ref{fig:CombGeo} a. was utilized where the moderator and reflector thicknesses were set to $r = 8$~cm and $t = 30$~cm, respectively.}
    \label{fig:T-neutrons}
\end{figure}

The Monte Carlo particle transport programs \textit{MCNP6} and \textit{Fluka} were utilized to confirm the simulation results which are reported in the NOVA ERA report \cite{Mauerhofer2017}. Two simple geometries were considered for comparison as shown in Figure \ref{fig:CombGeo}. The first geometry incorporates a polyethylene moderator which slows neutrons emanating from a Be target of thickness 0.7~mm. The target is mounted on an aluminum backing with a water cooling channel placed just behind it. This geometry was utilized in \textit{MCNP6} to obtain the spatial distribution of unperturbed thermal neutron flux as shown in Figure \ref{fig:T-neutrons}. These results reflect the same qualitative features which were presented in the NOVA ERA simulation with the maximum thermal neutron flux found $\sim$4.5~cm behind the target on the extended proton beam axis.

\begin{figure}[htbp!]
    \centering
    \includegraphics[width=0.9\textwidth]{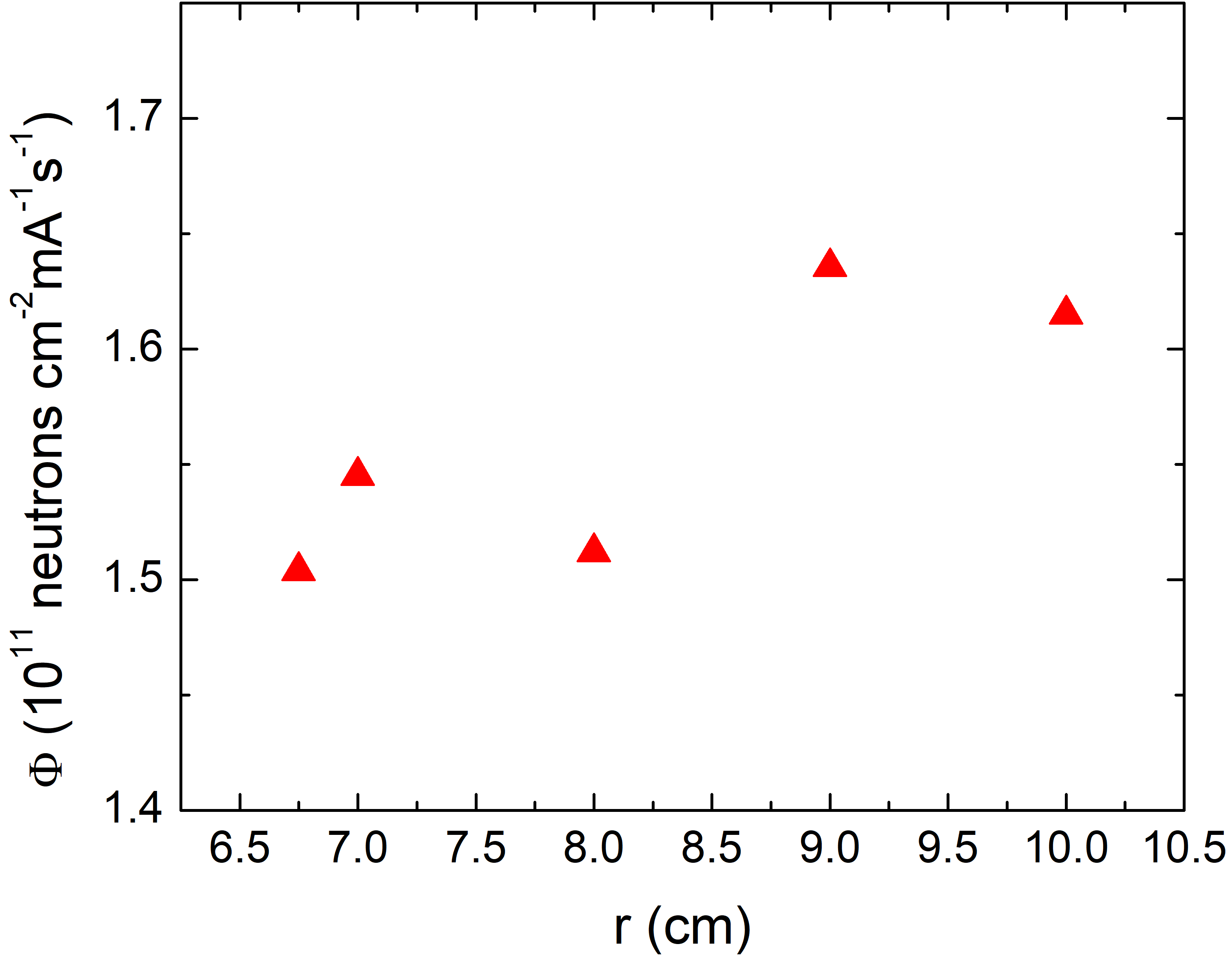}
    \caption{Variation of the thermal neutron flux emanating from the target as a function of moderator thickness, with the reflector thickness set to $t$ = 10~cm. Thermal neutrons were defined as neutrons possessing energies between 13.1 meV and 127 meV. These simulations were performed with the Monte Carlo transport code, \textit{Fluka}.}
    \label{fig:thneuflux}
\end{figure}

To make quantitative comparisons, simulations were performed with \textit{Fluka} by implementing the same geometry, except the moderator thickness $r$ was varied while the reflector thickness $t = 10$~cm was kept fixed. The thermal neutron flux was obtained by scoring the thermal neutron flux crossing the boundary between region $R1$ to region $R2$, where $R1$ is defined as a cone which opens in the forward direction, parallel to the $z$-axis (see Figure \ref{fig:CombGeo}). The angle subtended by the edge of the cone and the $z$-axis is 5$^\circ$. The results are presented in Figure \ref{fig:thneuflux} and can be compared directly with those presented in the NOVA ERA report. The trend showing the increase in the thermal neutron flux with increasing moderator size is generally reflected in our simulation results. The values which we obtained for the maximum neutron flux from our \textit{Fluka} simulations exceed the the maximum neutron flux given in the NOVA ERA report by a factor $\sim~6\%$.

Further quantitative comparisons were considered by employing the geometry shown in Figure \ref{fig:CombGeo}~b.). This is a simpler version of the geometry, known as \textit{Configuration A} in the NOVA ERA report, where only beam tubes C1 and T1 are considered. The thermal and cold neutron fluxes at the end of the cold beam tube (C1), just beyond the thickness of the lead reflector, were tallied while the thermal neutron flux was only tallied in the case of the thermal tube (T1). The results of the simulation are presented in Table \ref{Flux_comparisons}. The cold neutron yield obtained from \textit{Fluka} simulations in C1 is a factor 4 lower than those obtained with \textit{MCNP6}. The thermal neutron yield obtained from the \textit{Fluka} simulations are 3 and 10 times higher than those obtained with \textit{MCNP} in beam tubes C1 and T1.

It is known that \textit{Fluka} utilizes group-wise cross-sections for materials while \textit{MCNP6} makes use of point-wise cross-sections. The simulation results are influenced by this factor, but it is not well understood how the use of these two treatments of the cross-section influence simulation results in the low energy neutron regime. In addition, \textit{Fluka} is not optimized for neutron energies below the thermal regime. With these issues considered, we conclude that we can confirm that the results which the NOVA ERA group disclose in their conceptual design report are reasonable. The comparisons made in this paper highlight the importance of benchmarking \textit{Fluka} simulations with \textit{MCNP6} simulations in the low energy neutron regime. Future benchmarking simulations with \textit{Fluka} are being planned for direct comparison with simulations in \textit{MCNP} presented by Jeon \textit{et al} \cite{Jeon2020}.


\subsection{Target Design Challenges}

Key challenges that need to be addressed in target design are (1) the mitigation of hydrogen gas production in the target to preserve target longevity, (2) the efficient heat removal from the target and (3) target robustness (4) target exchange. The neutron flux at the final location of the scattering or BNCT sample depends on the desired neutron energy range, the moderator, reflector and most of all on the proton beam flux on the target. This calls for increasing the proton induced power density on the target surface to the maximum that can be accommodated. Target longevity is a desirable property for target design, especially in the context of BNCT applications, as easy target implementation and low variation in neutron beam quality and output determines the feasibility of designs. The issue of efficient heat removal also needs careful consideration in our target design. In the prototype stage, we intend to subject the target to modest beam powers of up to 2~kW in the first stage and higher as engineering allows. Target development studies have been performed for CANS projects worldwide and they provide important perspectives for our target design \cite{Mauerhofer2017,Yamagata2015,Rinckel2012,Kumada2019}.

First, we shall consider two approaches which are utilized to prevent ``blistering" of the target due to the formation of hydrogen gas within it. In the case of the early CANS project at the Low Energy Neutron Source in the USA, hydrogen blistering was found to destroy the target after 1 week of continuous operation of the 13~MeV proton beam at a beam power of 4~kW \cite{Rinckel2012}. To address this issue, some targets have been specifically designed so that the target thickness is smaller than the range of protons with a given incident energy. Therefore, the protons stop within the cooling water or in the backing material behind the target \cite{Rinckel2012,Mauerhofer2017}. To evaluate the range of protons in beryllium, the software package the \textit{Stopping Range of Ions in Matter (SRIM)} \cite{Ziegler2010} is typically employed, and the longitudinal straggling of the protons is subtracted from the range to provide a safe estimate for the beryllium thickness \cite{Jeon2020}. These findings can also be verified with simple simulations in \textit{Fluka} by locating the formation of the Bragg peak, associated with the stopping of protons, within the target. Another approach that has been employed in target design is the implementation of a layer of metal with high hydrogen diffusivity. For the RANS project, vanadium was found to outperform other candidate backing materials with high hydrogen diffusivity including palladium, tantalum and niobium \cite{Yamagata2015}.

A further development of this approach has been recently proposed by Kurihara \textit{et al.} \cite{Kurihara2020}, wherein a thin layer of Be was successfully bonded via diffusion bonding to a two-layer mount containing a layer of palladium and a layer of copper. The beryllium layer is sufficiently thin to ensure that protons do not stop within it. The second layer of the target is made of palladium, which has a much higher hydrogen diffusivity than vanadium. The third layer is comprised of copper, which has excellent thermal conductivity to aid in efficient heat removal \cite{Kurihara2020}. This design is of particular interest for implementation in our project because the iBNCT and KEK groups have shown that this target design is capable of withstanding exposure of up to 6,000 Coulombs of proton charge or more \cite{Kumada2019} or 1670~mA/hours (or 835 patient-hours for a treatment proton current of 2~mA). In the context of BNCT applications this exposure is equivalent to 1500 patients, assuming 4 C is required for treating each patient \cite{Kumada2019} (0.55~hours at 2~mA). 

TRIUMF operates several dedicated test facilities for the development of target materials and target designs. The Converter Test Facility that is permanently installed at TRIUMF's electron linac allows to subject potential target materials or material composites to surface power densities in excess of 20~W/mm$^2$ at longitudinal beam stopping profiles that are similar to a 10 MeV proton beam. In parallel to themo-mechanical validations using the electron beam, TRIUMF's TR13 cyclotron can deliver proton beams of 13 MeV and 100~$\mu$A which can be used to induce radiation damage and material blistering / swelling. Together with in-house characterization capabilities of activated material samples, including electron microscopy, this will allow systematic development of a capable target material for PC-CANS. 

\subsection{Neutron beamline design}
The PC-CANS will supply neutrons to only two neutron scattering instruments, a SANS and a general purpose diffractometer. Consequently, the layout of beam tubes will be significantly different from the arrangement proposed for the NOVA ERA.  Two tangential beam tubes are currently considered as shown in Fig. \ref{fig:CANS_schematic}. Whereas SANS is permanently on the cold source, the diffractometer can choose to receive neutrons from the cold source or from a thermal source if the cavity at the start of its beam tube is filled with water.

\section{End Stations}
\subsection{Overview}
Many societal issues in energy, the environment, computing, manufacturing, and health call for technological solutions based on high-performing materials. In industry and at universities alike, many research teams aim to enhance the performance of materials and create innovative technologies based on new materials. Inevitably, some of these teams encounter problems that can only be solved using neutron beams, as neutrons interact with materials in ways fundamentally different from complementary tools such as beams of light, x‑rays, and electrons. 

\subsection{SANS End Station}
The workhorse neutron scattering instrument at PC-CANS is proposed to be a Small-Angle Neutron Scattering (SANS) facility. Factors in this selection include local research interests at the University of Windsor, as well as the Canadian and international demand for SANS. SANS has a broad multi-disciplinary appeal. Diffracted intensity measurement at low scattering angles can tolerate large uncertainty in neutron wavelength, enabling SANS at a relatively low intensity pulse neutron source. The performance of the time-of-flight SANS at PC-CANS can be directly compared with the SANS facility at the Low-Energy Neutron Source (an early CANS prototype in Indiana)~\cite{Baxter2016} and at the McMaster Nuclear Reactor.

\begin{figure}[htbp!]
    \centering
    \includegraphics[width=0.95\textwidth]{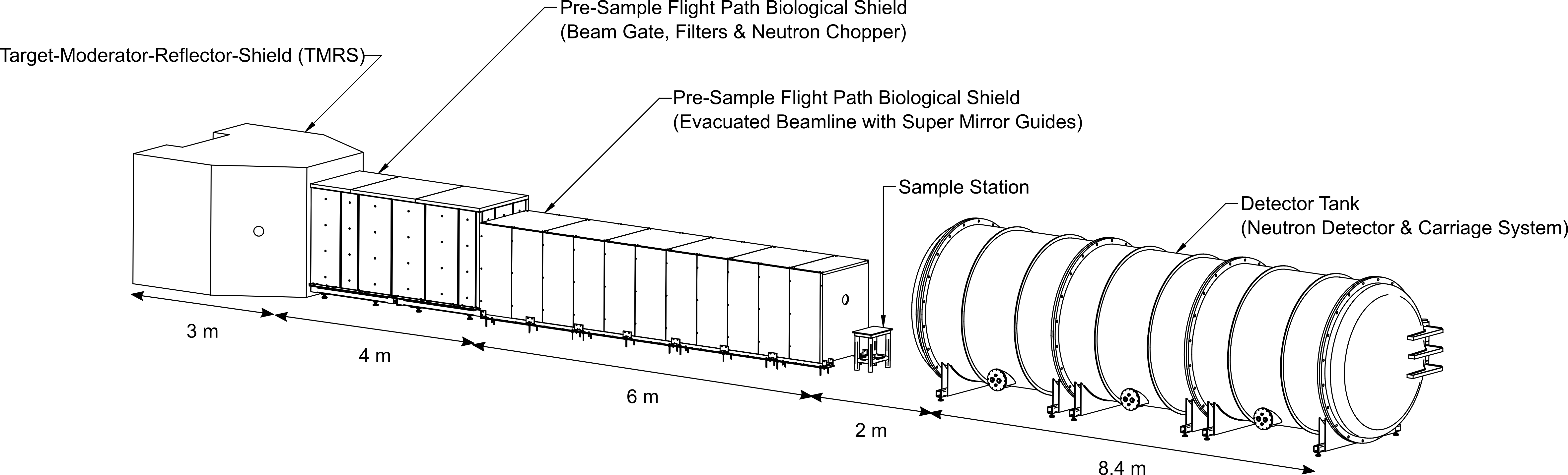}
    \caption{Schematic of a proposed small angle neutron scattering instrument for the PC-CANS.}
    \label{fig:SANS}
\end{figure}

The SANS instrument at PC-CANS is envisioned be to 20 m long, roughly 10 m from the cold-source to the sample position and 10 m further to the maximum distance to the 1 m$^2$ detector. Being ~40\% longer that the SANS planned for the NOVA ERA, this instrument is designed to take advantage of the five times higher neutron flux at the source.  The general layout is shown in Figure \ref{fig:SANS} while the expected performance is as listed in Table~\ref{Tab:SANS}.

\subsection{Multipurpose – diffraction/imaging station}

We are proposing to include a modular, flexible, and versatile instrument at PC-CANS facility to perform neutron diffraction, residual stress determination, and imaging (diffraction imaging multipurpose station, DIMS) that allows each of these measurements individually or simultaneously. The modular aspect of the proposed instrument makes it easy to change the setup for performing these measurements individually or combined when needed to enable, for example, simultaneous diffraction and imaging data acquisition.

\begin{figure}[htbp!]
    \centering
    \includegraphics[width=0.95\textwidth]{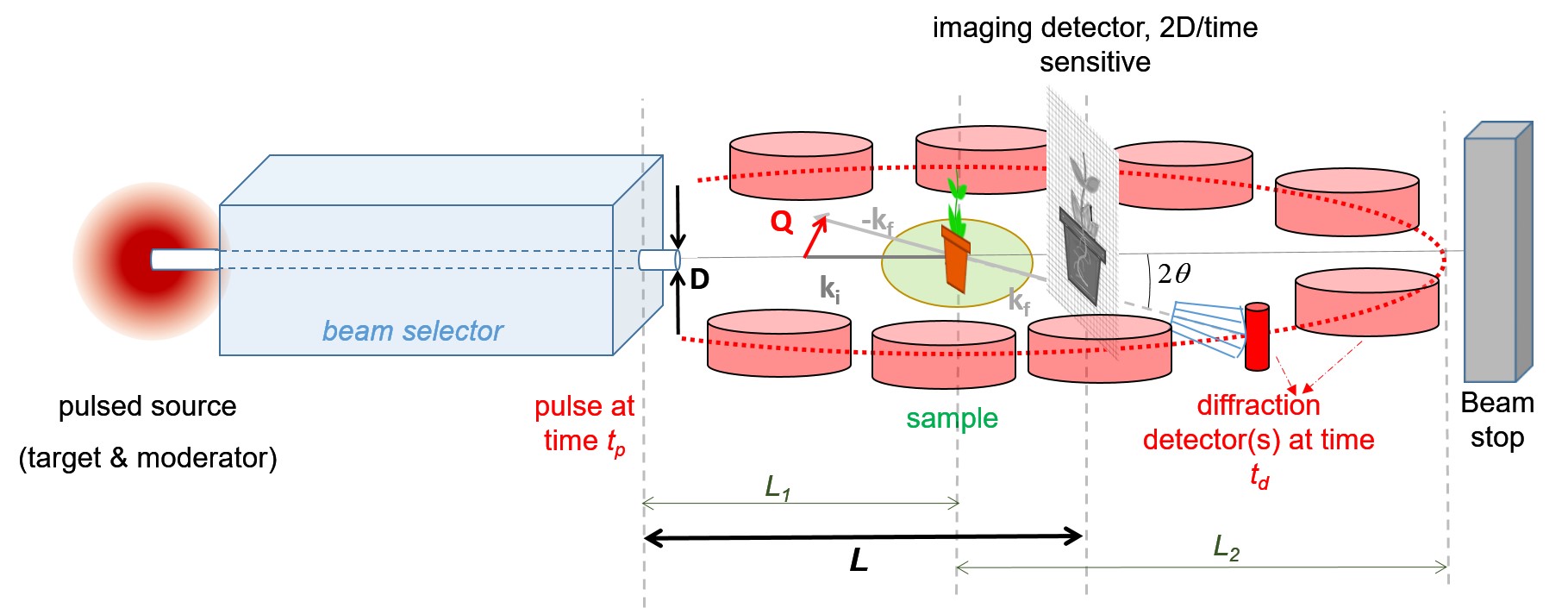}
    \caption{Schematic of a proposed combined diffraction and imaging system for the PC-CANS.}
    \label{fig:diffr}
\end{figure}

A schematic representation of DIMS is shown in Figure \ref{fig:diffr}. The neutron beam extracted from the target and moderator assembly is optimized for the measurements using the components of the beam selector (such as velocity selector, choppers, filters such as sapphire, shielding, focusing guides, and masks to control the beam size). The spectra of the exit beam from the selector is monitored for the required normalization of counts at the detectors. To enhance the flux sample, a rotary stage is located at a short distance from the exit of the beam selector. A diffraction detector (or blocks of detectors) and associated radial collimators are used for measuring diffraction intensity as a function of scattering angle. A two-dimensional time-sensitive transmission imaging detector will be located close to the sample to increase the image sharpness. A beam stop is located at the straight-through position at the far side of the instrument to block the main beam outside of the experimental area.

Considering that the PC-CANS brightness would be five times larger than the NOVA-ERA source,~\cite{Mauerhofer2017} we expect the averaged thermal neutron flux of the order of at least 2x10$^4$ n/cm$^{2}$/s for DIMS which is of similar order of magnitude as at the RANS facility. It is expected additional gain and enhanced performance can be achieved with optimization of the beam optics and components of the instrument (distances, focusing neutron guides with divergence changers, shielding, etc.). 

\section{Applications} 
\subsection{Boron Neutron Capture Therapy}
An estimated 226,000 Canadians will have been diagnosed with cancer in 2020, and about 37\% of these patients will die from it~\cite{Brenner2020}. BNCT is a radiation therapy~\cite{Suzuki2020,Kiyanagi2019,Barth2018,MIYATAKE2016, Nedunchezhian2016,Moss2014,Suzuki2014,Kato2020,Koivunoro2019,Barth2012} that could be effective in many of these cases, but is most urgently needed for recurrent head and neck cancers and glioblastoma multiforme (GBM), for which conventional treatments usually fail. For example, over 80\% of the projected 3000 Canadians who will be diagnosed in 2020 with cancer in the brain or central nervous system will die from it ~\cite{Brenner2020}. 

To ensure relevance of pre-clinical BNCT research at the PC-CANS, we will strive to reproduce key parameters of the neutron beam used for clinical BNCT at the Southern Tohoku BNCT Research Center (STBRC) in Japan. These parameters include purity, energy spectrum, spatial distribution, and size ~\cite{Kiyanagi2019,Tanaka2009,Tanaka2011,Agency2001}, but it will not compete on brightness. Rather, the PC-CANS will be used demonstrate pre-clinical capabilities that could justify investment in a clinical BNCT machine at a later stage. Since our CANS design will be scalable, its brightness could be increased in a future phase of the research through operation in dedicated mode and boosting the power of the linac--a strategy currently being employed by the Ibaraki BNCT project in Japan~\cite{Kiyanagi2019}.

\subsection{PET Isotope Production}

Cyclotrons are the most common accelerator for medical applications; nevertheless, linear accelerators are equally capable and can be cost-effective when teamed with other applications as in PC-CANS. The utility of a linac is its ability to host multiple end-stations at different energies~\cite{Oliver:2017nxg}. The accelerator for our CANS prototype will be used to produce the [18F]-fluorodeoxyglucose (FDG) for positron emission tomography (PET) scans. Among the breadth of novel PET tracers being developed, 96\% of PET scans still require a radiopharmaceutical [18F]-fluorodeoxyglucose, known as FDG~\cite{Achmad2017}. The isotope fluorine-18 decays rapidly, with a half-life of 110 minutes. Such a short half-life makes shipping the FDG long distances expensive and often impractical. PET scanners that source their FDG from long distances are limited in the number of patients they can serve, and if the full local demand cannot be met, patients must travel elsewhere or accept much longer wait times, which may be detrimental to their health. Given the proposed host location for the CANS prototype is located in a city with a PET scanner and no local FDG supply, the ability to produce the essential imaging tracers at the CANS facility will be valued by the local health system. As such, accelerator parameters were defined with this application in mind. Furthermore, the infrastructure could be used to produce other medical isotopes for PET scanning that cannot be imported at all because they have even shorter half-lives: $^{15}$O (2 min), $^{13}$N (10 min), and $^{11}$C (20 min).

\section{Summary}

We have conceptualized a prototype CANS facility, PC-CANS, based on linac technology that could lead to a brighter CANS facility that is well suited to Canada's needs for a new neutron source. PC-CANS will allow competitive rates for neutron science, exceeding the NOVA ERA proposal by a factor of five, while simultaneously producing F-18 for PET and protons to a dedicated BNCT research and development facility. As target technology develops, the linac will allow dedicated operation at higher intensities for either Neutron Science or BNCT end stations up to an average power of 10~kW and peak power of 200~kW. These latter values would exceed the yield presented in the NOVA proposal by a factor of 25. A summary of the staging scenarios and production at the three stations is given in Table~\ref{Tab_Fsummary}.

The PC-CANS prototype will be the first CANS in Canada and the first for BNCT in North America. Our efforts will position Canada among the first countries to realize BNCT for clinical use by building the success of contemporary facilities around the world. Impact on the domestic healthcare will be further realized through our production of PET radioisotopes. CANS technology offers a new paradigm in which Canada could satisfy many of its research needs for neutron beams with a relatively modest investment while making these irreplaceable tools much more accessible to Canadian researchers. Lowering access barriers to these scientific tools will also promote equity and diversity in science and in training of the next generation of neutron users. The impact of better access will be to accelerate Canadian research and innovation for which neutrons are required. 

\section*{Acknowledgments}
The authors thank Dr. Thomas Gutberlet for fruitful discussions. The authors acknowledge the support from the New Frontiers in Research Fund-Exploration grant (NFRFE-2018-00183) and the University of Windsor Office of Research and Innovation.




\nocite{*} 
\bibliographystyle{ios1}           
\bibliography{bibliography}        
\onecolumn
%
\newpage

\newpage

\newpage

\newpage

\newpage
\newpage

\newpage

\newpage

\begin{table}[htbp!]
\caption{Simplified groupings of neutron sources by brightness (i.e. performance of diffraction instruments), comparing costs of conventional sources with CANS. Dagger ($\dagger$) indicates facilities that are envisioned.}
\begin{tabular}{lccc}
\multicolumn{2}{c}{\textbf{Relative Performance}} & \multicolumn{1}{c}{\textbf{Conventional Sources}} & \multicolumn{1}{c}{\textbf{CANS}}                  \\ \hline \hline
High   & 5-10+  & SNS (\$2B); ESS (\$3B)     & ---  \\ \hline
Medium & 1 & ISIS (\$850M); NRU ($\>$\$500M)     & Canada-scale facility$^\dagger$ (\$100-200M)     \\ \hline
Medium-Low  & 1/5  & MNR (\textgreater{}\$100M)   & \textbf{PC-CANS$^\dagger$ (\$10-12M + 2 instruments)} \\ \hline
Low  & 1/25 &  ---   & NOVA ERA$^\dagger$ (\$6M + 6 instruments); LENS; RANS
\end{tabular}
\label{Tab_sources}
\end{table}

\newpage
\begin{table}[htbp!]
\caption{Summary of PC-CANS requirements}
\begin{tabular}{l|c|c|c|c|c|c}
Station                & Energy (MeV) & I$_{ave}$ ($\mu$A) & Duty Factor  & P$_{ave}$ (kW) & I$_{peak}$ (mA) & P$_{peak}$ (kW) \\ \hline
NOVA ERA                   & 10           & 40                & 4\%          & 0.4            & 1                   & 10              \\
HBS                    & 70           & 4300              & 4.30\%       & 301            & 100                 & 7000            \\
\hline
Neutron                & 10           & 200               & 5\%          & 2              & 4                   & 40              \\
$^{18}$F                   & 10           & 100               & 5\%          & 1              & 2                   & 20              \\
BNCT                   & 10           & 200               & 5\%          & 2              & 4                   & 40              \\ \hline
\textbf{Target Totals} & \textbf{}    & \textbf{500}      & \textbf{}    & \textbf{5}     & \textbf{10}         & \textbf{100}    \\
\hline \hline
\textbf{Linac Totals}  & \textbf{10}  & \textbf{1000}     & \textbf{5\%} & \textbf{10}    & \textbf{20}         & \textbf{200}   
\end{tabular}
\label{Tab_summary}
\end{table}

\newpage

\begin{table}[htbp!]
\caption{\label{Flux_comparisons}Results from simulations to be compared with NOVA ERA studies are presented in this table. Energy ranges are defined as follows: 13.1 meV to 127 meV for thermal neutrons and less than 13.1 meV for cold neutrons. The calculated neutron fluxes (cm$^{-2}$s$^{-1}$mA$^{-1}$) are defined in the forward direction with an angle 0.5$^\circ$ with respect to the normal of the surface.}
\begin{tabular}{c|c|cc}
\hline
\multicolumn{4}{c}{\textbf{Neutron Yields From Channels In Configuration A}} \\ \hline
\multirow{2}{*}{\textbf{Energy Range}} & \multirow{2}{*}{\textbf{Source}}  & \textbf{C1} & \textbf{T1}  \\
 &  & \multicolumn{2}{c}{(\textit{\textbf{cm$^{-2}$s$^{-1}$mA$^{-1}$)}}}  \\
\hline
\multirow{3}{*}{Cold neutrons}      & This work  & 3.41~$\cdot$~10$^{5}$ & $-$ \\
                                    & NOVA ERA  & 1.12~$\cdot$~10$^{6}$ & $-$ \\
                                    & Ratio  & 0.3  & $-$ \\
                                    \hline
\multirow{3}{*}{Thermal neutrons}  & This work  & 4.73~$\cdot$~10$^{6}$  &                                                  8.46~$\cdot$~10$^{6}$ \\
                                   & NOVA ERA  & 4.76~$\cdot$~10$^{5}$ & 2.93~$\cdot$~10$^{6}$ \\
                                   & Ratio  & 9.9 & 2.9 \\
                                   \hline
\end{tabular}
\end{table}

\newpage

\begin{table}[htbp!]
\caption{Collimation and performance of the proposed small angle neutron scattering instrument for the PC-CANS}
\begin{tabular}{c|c|c|c|c|c}
\textbf{Collimation (m)} & \textbf{Sample to detector (m)} & \textbf{Diameter of S1 (mm)} & \textbf{Diameter of S2 (mm)} & \textbf{l band (\AA)} & \textbf{Q$_{min}$  (\AA$^{-1}$)} \\ \hline
1& 1& 20& 10& 4 – 10              & 0.019                \\ \hline
2& 2& 20& 10& 4 – 9.5             & 0.010                \\ \hline
4& 4& 20& \textbf{10}& 4 – 8.8& 0.0054               \\ \hline
6& 6& 20& 10& 4 – 8.2             & 0.0038               \\ \hline
10& 10& 20& 10& 4 – 7.5& 0.0025              
\end{tabular}
\label{Tab:SANS}
\end{table}

\newpage \begin{table}[htbp!]
 \caption{Expected Performance for 10 MeV protons}
\begin{tabular}{|ll|c|c|c|c|}
\hline
\multicolumn{2}{|c|}{\textbf{Applications}}       & \multicolumn{4}{c|}{\textbf{I$_{ave}$/I$_{peak}$}}   \\
\multicolumn{2}{|c|}{\textbf{}}                   &  0.1/2 & 0.2/4   & 0.5/10   & 1/20      \\ \hline
Neutron Science & Cold (n/cm$^2$/s)$^a$       &   ---    & 8$\cdot~10^5$/1.6$\cdot~10^7$   & 2$\cdot~10^6$/4$\cdot~10^7$  & 4$\cdot~10^6$/8$\cdot~10^7$   \\
                & Thermal (n/cm$^2$/s)$^a$    &   ---    & 1.2$\cdot~10^6$/2.4$\cdot~10^7$ & 3$\cdot~10^6$/6$\cdot~10^7$  & 6$\cdot~10^6$/1.2$\cdot~10^8$ \\ \hline
BNCT            & Epithermal (n/cm$^2$/s)$^b$ &   ---    & 1.0$\cdot~10^8$    & 2.5$\cdot~10^8$ & 5.0$\cdot~10^8$  \\ \hline
PET             & 18F (GBq)$^c$               & 240   &     ---        &     ---     &     ---     \\ \hline
\multicolumn{6}{l}{$^a$ Yield at surface of reflector with +0.5 deg} \\ 
\multicolumn{6}{l}{$^b$ Yield from MgF$_2$ (ref \cite{Kiyanagi2019}} \\ 
\multicolumn{6}{l}{$^c$ Saturated yield} \\ 
\end{tabular}
\label{Tab_Fsummary}
\end{table}

\end{document}